\documentclass[%
 reprint,
 superscriptaddress,
 amsmath,amssymb,
 prl,
]{revtex4-1}
\usepackage{graphicx}
\usepackage{dcolumn}
\usepackage{bm}
\usepackage{float}
\usepackage{siunitx}
\usepackage{color}
\usepackage{amsmath}
\usepackage{mathtools}
\usepackage[colorlinks=true, linkcolor=black,urlcolor=blue,citecolor=blue]{hyperref}

\begin{document}

\preprint{APS/123-QED}

\title{Generating Spatially Entangled Itinerant Photons with Waveguide Quantum Electrodynamics}

\author{Bharath Kannan}
\email{bkannan@mit.edu}
 \affiliation{Research Laboratory of Electronics, Massachusetts Institute of Technology, Cambridge, MA 02139, USA}
\affiliation{Department of Electrical Engineering and Computer Science, Massachusetts Institute of Technology, Cambridge, MA 02139, USA}

\author{Daniel L. Campbell}
 \affiliation{Research Laboratory of Electronics, Massachusetts Institute of Technology, Cambridge, MA 02139, USA}

\author{Francisca Vasconcelos}
 \affiliation{Research Laboratory of Electronics, Massachusetts Institute of Technology, Cambridge, MA 02139, USA}
\affiliation{Department of Electrical Engineering and Computer Science, Massachusetts Institute of Technology, Cambridge, MA 02139, USA}

\author{Roni Winik}
 \affiliation{Research Laboratory of Electronics, Massachusetts Institute of Technology, Cambridge, MA 02139, USA}
 
 \author{David Kim}
\affiliation{MIT Lincoln Laboratory, 244 Wood Street, Lexington, MA 02420, USA}

\author{Morten Kjaergaard}
 \affiliation{Research Laboratory of Electronics, Massachusetts Institute of Technology, Cambridge, MA 02139, USA}

\author{Philip Krantz}
\altaffiliation[Present address: ]{Wallenberg Centre for Quantum Technology, Department of Microtechnology and Nanoscience, Chalmers University of Technology, 412 96 Gothenburg, Sweden}
 \affiliation{Research Laboratory of Electronics, Massachusetts Institute of Technology, Cambridge, MA 02139, USA}

\author{Alexander Melville}
\affiliation{MIT Lincoln Laboratory, 244 Wood Street, Lexington, MA 02420, USA}

\author{Bethany M. Niedzielski}
\affiliation{MIT Lincoln Laboratory, 244 Wood Street, Lexington, MA 02420, USA}

\author{Jonilyn Yoder}
\affiliation{MIT Lincoln Laboratory, 244 Wood Street, Lexington, MA 02420, USA}

\author{Terry P. Orlando}
 \affiliation{Research Laboratory of Electronics, Massachusetts Institute of Technology, Cambridge, MA 02139, USA}
\affiliation{Department of Electrical Engineering and Computer Science, Massachusetts Institute of Technology, Cambridge, MA 02139, USA}

\author{Simon Gustavsson}
 \affiliation{Research Laboratory of Electronics, Massachusetts Institute of Technology, Cambridge, MA 02139, USA}

\author{William D. Oliver}
 \affiliation{Research Laboratory of Electronics, Massachusetts Institute of Technology, Cambridge, MA 02139, USA}
 \affiliation{Department of Electrical Engineering and Computer Science, Massachusetts Institute of Technology, Cambridge, MA 02139, USA}
\affiliation{MIT Lincoln Laboratory, 244 Wood Street, Lexington, MA 02420, USA}
\affiliation{Department of Physics, Massachusetts Institute of Technology, Cambridge, MA 02139, USA}





\begin{abstract}
 Realizing a fully connected network of quantum processors requires the ability to distribute quantum entanglement. For distant processing nodes, this can be achieved by generating, routing, and capturing spatially entangled itinerant photons. In this work, we demonstrate the deterministic generation of such photons using superconducting transmon qubits that are directly coupled to a waveguide. In particular, we generate two-photon N00N states and show that the state and spatial entanglement of the emitted photons are tunable via the qubit frequencies. Using quadrature amplitude detection, we reconstruct the moments and correlations of the photonic modes and demonstrate state preparation fidelities of $84\%$. Our results provide a path towards realizing quantum communication and teleportation protocols using itinerant photons generated by quantum interference within a waveguide quantum electrodynamics architecture.
\end{abstract}
\maketitle

%
%

\begin{figure*}
    \centering
    \includegraphics[width=\textwidth]{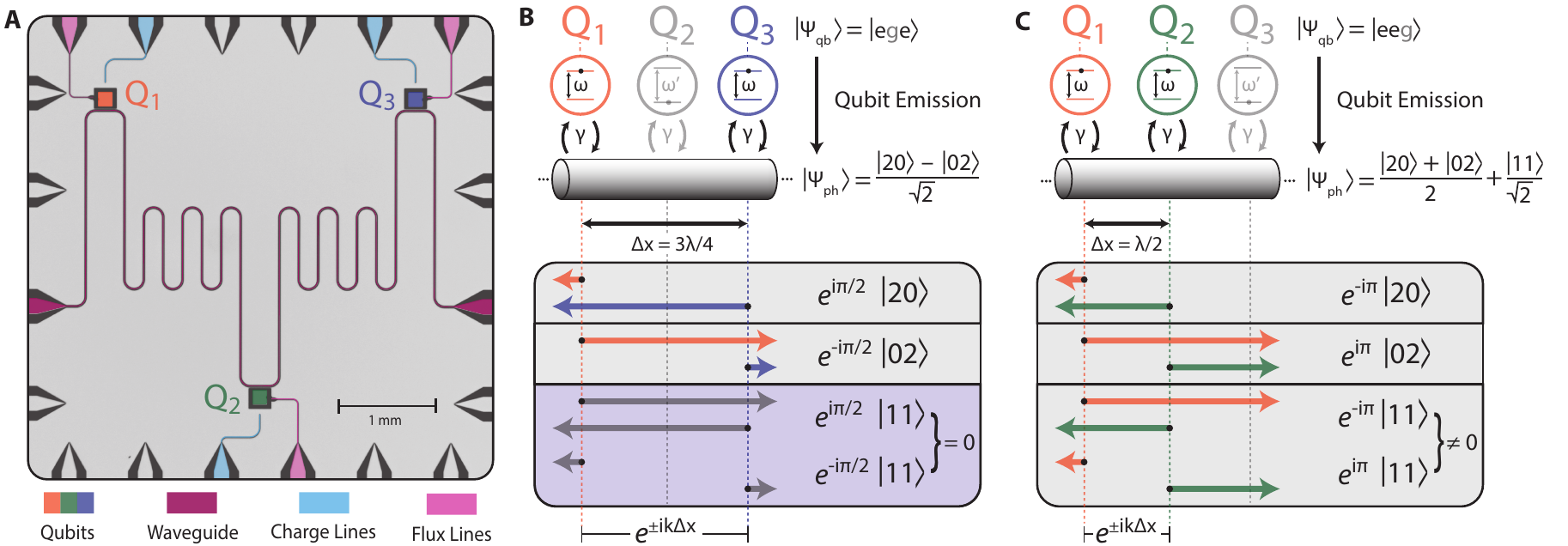}
    \caption{\textbf{Generating spatially correlated itinerant photons in wQED.}\ (\textbf{A}) A false-colored micrograph of the device. The device consists of three independently flux-tunable transmon qubits that are capacitively coupled to a common waveguide. (\textbf{B}) Schematic diagram of three qubits that are coupled to a common waveguide with equal strength $\gamma$. Qubits $Q_1$ and $Q_3$ are initially excited and placed on resonance at $\omega/2\pi = \SI{4.85}{GHz}$ such that their spatial separation along the waveguide is $\Delta x = 3\lambda/4$. Qubit $Q_2$ is detuned far away $|\omega' - \omega| \gg \gamma$ such that it can be ignored and is left in the ground state. The four possible coherent pathways for the photons emitted by the qubits into the left and right travelling modes of the waveguide are shown below. The state of the emitted photons is a two-photon N00N state due to destructive interference between the single-photon pathways $|11\rangle$. (\textbf{C}) The same setup as (B) except $Q_1$ and $Q_2$ are now placed on resonance $\omega/2\pi = \SI{6.45}{GHz}$ such that $\Delta x = \lambda/2$ and $Q_3$ is now detuned far away. The $|11\rangle$ states constructively interfere for this choice of $\Delta x$. }
    \label{fig:1}
\end{figure*}

\begin{figure}[h]
    \centering
    \includegraphics[width=3.3in]{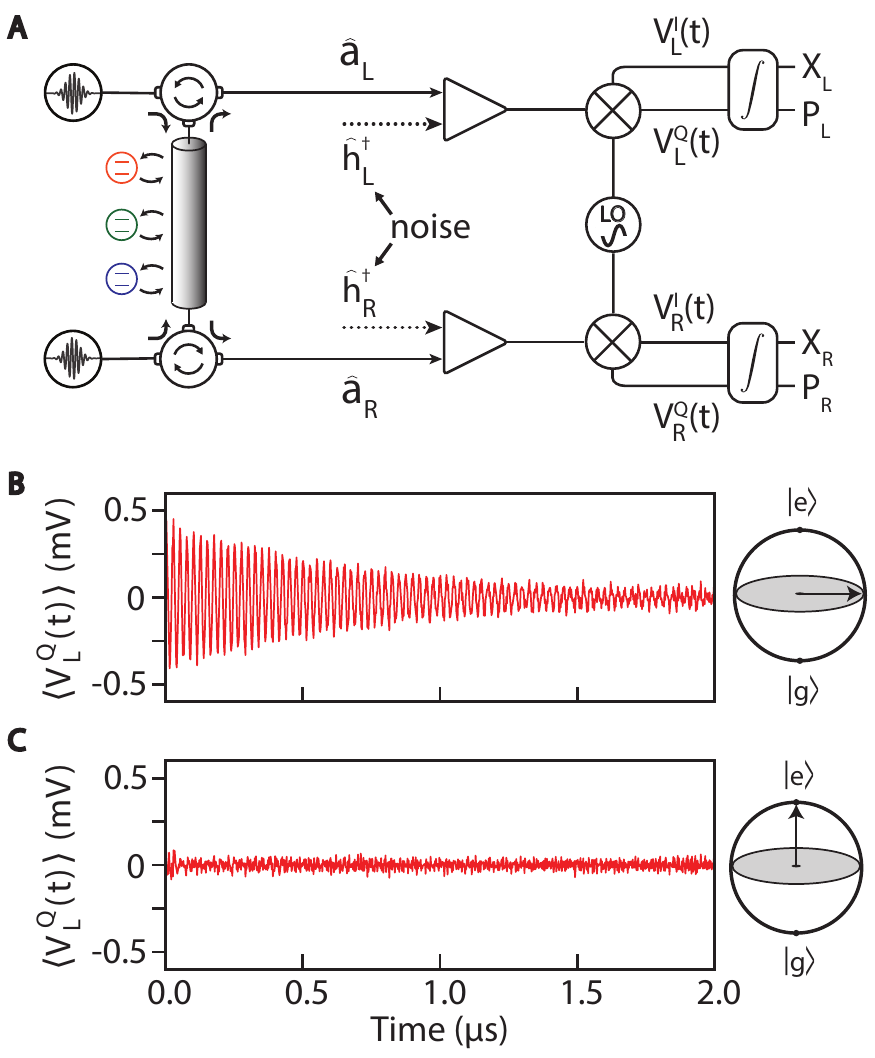}
    \caption{\textbf{Measurement setup and procedure.}\ (\textbf{A}) A schematic setup of the dual-sided control and measurement chain. The signal from the photons emitted by the qubits is amplified and downconverted to an intermediary frequency $f_d$ before digitization. The digitized signal is then further demodulated and integrated using custom FPGA code to obtain a pair of complex numbers $S_\textrm{L} = X_\textrm{L}+iP_\textrm{L}$ and $S_\textrm{R} = X_\textrm{R}+iP_\textrm{R}$. Single-shot measurements of these values are then binned into a histogram to construct a 4D probability distribution. The mode of interest, $\hat{a}_\textrm{L(R)}$, and noise mode, $\hat{h}_\textrm{L(R)}^\dag$, of the left (right) measurement chain are indicated directly prior to amplification. (\textbf{B}) A representative time trace of the digitized and averaged voltage from the emission of a single qubit initialized to $(|g\rangle + |e\rangle)/\sqrt{2}$. The exponential temporal envelope of the emission is superimposed with oscillations at the downconverted frequency $f_d = \SI{40}{MHz}$. (\textbf{C}) The voltage from the emission of a qubit initialized to $|e\rangle$. The photon is emitted with a random phase such that the voltage averages to zero. }
    \label{fig:2}
\end{figure}

Modular architectures of quantum computing hardware have recently been proposed as an approach to realizing robust large-scale quantum information processing \cite{Cirac1999,Kimble2008,Monroe2014,Jiang2007}. However, such architectures rely on a means to coherently transfer quantum information between individual, and generally non-local, processing nodes. Spatially entangled itinerant photons can be used to achieve this by efficiently distributing entanglement throughout a quantum network. Conventional approaches for generating such photons in optical systems typically utilize spontaneous parametric down conversion in conjunction with arrays of beamsplitters \cite{Afek2010} and photodetectors for post-selection \cite{Kok2002,Walther2004}. However, the stochastic nature of these approaches limit their utility in quantum information processing applications. 

Recent progress with superconducting circuits has established a path towards realizing a universal quantum node that is capable of storing, communicating, and processing quantum information \cite{Saito2013,Kurpiers2018,Leung2019,Barends2014,Arute2019}. These works often invoke a cavity quantum electrodynamics (cQED) architecture, where cavities protect qubits from decoherence within a node, enabling the high-fidelity control required to generate arbitrary quantum states. To link distant nodes, this quantum information must propagate along a bus comprised of a continuum (or quasi-continuum) of modes. To this end, we strongly couple qubits to a waveguide such that the excitations stored in the qubits are rapidly released as itinerant photons. Such a system is described by waveguide quantum electrodynamics (wQED). Entering the strong coupling regime in wQED enables qubits to serve as high-quality quantum emitters \cite{Abdumalikov2011}. More generally, superconducting circuits have been used to produce a wide variety of non-classical itinerant photons from classical drives \cite{Hoi2012,Forn-Diaz2017,Gonzalez-Tudela2015,Pfaff2017,Gasparinetti2017}, such as those with correlations and entanglement in frequency \cite{Gasparinetti2017}.

Here, we demonstrate that the indistinguishability and quantum interference between photons directly emitted from multiple sources into a waveguide can deterministically generate spatially entangled itinerant photons. In particular, we generate two-photon N00N states $|\psi_\textrm{ph}\rangle = (|20\rangle - |02\rangle)/\sqrt{2}$, where the state $|n_\textrm{L}n_\textrm{R}\rangle$ denotes the number of photons in the left and right propagating modes of the waveguide, respectively. More generally, we show that our device can generate itinerant photons with states of the form $|\psi_\textrm{ph}\rangle = a|20\rangle + b|02\rangle + c|11\rangle$, where $a$, $b$, and $c$ are complex coefficients that are set by the effective qubit spatial separation $\Delta x$.

The test device consists of three flux-tunable transmon qubits \cite{Koch2007} that are capacitively coupled to a common $\SI{50}{\Omega}$ transmission line (an electromagnetic coplanar waveguide), as shown in Fig.~\ref{fig:1}\textbf{A}. The configurations we consider involve two qubits, used as photonic emitters, that are spatially separated by $\Delta x = 3\lambda/4$ and $\Delta x = \lambda/2$. The effective spacing is controlled by the qubit frequencies $\omega$ \cite{VanLoo2013} via the corresponding wavelength $\lambda = 2\pi v/\omega$, where $v$ is the speed of light in the waveguide. Setting the transition frequencies of qubits $Q_1$ and $Q_3$ to $\omega/2\pi = \SI{4.85}{GHz}$ corresponds to a spacing of $\Delta x = 3\lambda/4$ between emitters. The frequency of the central qubit $Q_2$ is detuned hundreds of MHz such that it can be ignored. In this configuration, the qubits are coupled to the coplanar waveguide with a coupling strength of $\gamma/2\pi = \SI{0.53}{MHz}$. Alternatively, to realize a spacing of $\Delta x = \lambda/2$ between emitters, the frequencies of $Q_1$ and $Q_2$ are set to $\omega/2\pi = \SI{6.45}{GHz}$, where the qubit-waveguide coupling strength is $\gamma/2\pi = \SI{0.95}{MHz}$, while sufficiently detuning $Q_3$ that it may be ignored.

The Hamiltonian of the system is \cite{Lalumiere2013}

\begin{align}{}
\label{eq:hamiltonian}
&\hat{H}=\int_{0}^{\infty} d \omega \hbar \omega\left[\hat{a}_{\mathrm{L}}^{\dagger}(\omega) \hat{a}_{\mathrm{L}}(\omega)+\hat{a}_{\mathrm{R}}^{\dagger}(\omega) \hat{a}_{\mathrm{R}}(\omega)\right]+
\sum_{j} \frac{\hbar \omega_{j}}{2} \hat{\sigma}_{z}^{(j)} \nonumber \\
&-i \sum_{j} \int_{0}^{\infty} d \omega \hbar g_{j}(\omega)\hat{\sigma}_{x}^{(j)}\left[\hat{a}_{\mathrm{L}}^{\dagger}(\omega) e^{\frac{-i\omega x_j}{v}}+\hat{a}_{\mathrm{R}}^{\dagger}(\omega) e^{\frac{i\omega x_j}{v}}+ \textrm{h.c.}\right]
\end{align}
where $\hat{a}_\textrm{L(R)}^{\dag}(\omega)$ and $\hat{a}_\textrm{L(R)}(\omega)$ are the creation and annihilation operators for left (right) propagating photons with frequency $\omega$, $x_j$ is the position of the  $j^\textrm{th}$ qubit, and $\hat{\sigma}^{(j)}_x$ and $\hat{\sigma}^{(j)}_z$ are the qubit X and Z Pauli operators. The coupling strength $g_{j}(\omega)$ determines the physical qubit-waveguide coupling rate $\gamma(\omega_j) = 4\pi g_j(\omega_j)^2 D(\omega_j)$, where $D(\omega)$ is the density of photonic modes in the waveguide. The qubits couple to the transmission line with equal strength when placed on resonance with each other. 

When the propagation time for photons between the qubits is small relative to the timescale $\gamma^{-1}$ of the qubit emission, this system can be simulated for arbitrary initial conditions and spacings by integrating a master equation derived from the Hamiltonian in Eq. \ref{eq:hamiltonian} and applying input-output theory \cite{Lalumiere2013}. The input-output relations that provide the dynamics of the photons emitted into the left- and right-propagating modes at the qubit frequencies are
\begin{equation}
\label{eq:inout}
\begin{split}
        &\hat{a}_\textrm{L}^{}(t) = \hat{a}_\textrm{L}^\textrm{in}(t) + \sqrt{\frac{\gamma}{2}} (\hat{\sigma}_-^{(1)} + \hat{\sigma}_-^{(2)}e^{\frac{-i\omega \Delta x}{v}}) \\
        &\hat{a}_\textrm{R}^{}(t) = \hat{a}_\textrm{R}^\textrm{in}(t) + \sqrt{\frac{\gamma}{2}} (\hat{\sigma}_-^{(1)} + \hat{\sigma}_-^{(2)}e^{\frac{i\omega \Delta x}{v}}),
\end{split}
\end{equation}
where $\hat{a}_\textrm{L/R}^\textrm{in}(t)$ are the incoming field operators at time $t$, and are taken to be in the vacuum state.

The two resonant qubits in each spacing configuration are initialized to their excited states while the detuned qubit is left in the ground state. Under these conditions, the final (un-normalized) state of the photons emitted by the excited qubits is given by 
\begin{align}
\label{eq:state}
    |\psi_\textrm{ph}\rangle \otimes |gg\rangle = \prod_j^2\   (\hat{a}_\textrm{L}^{\dagger}e^{\frac{i\omega x_j}{v}}+\hat{a}_\textrm{R}^{\dagger}e^{\frac{-i\omega x_j}{v}})  |00\rangle \otimes \hat{\sigma}^{(j)}_-|ee\rangle,
\end{align}
where the photonic modes $\hat{a}_{\textrm{L(R)}}$ have been integrated over temporally, and the index $j$ is multiplied over the two active qubits that are prepared in the state $|ee\rangle$ ($Q_1$,$Q_3$ for $\Delta x=3\lambda/4$ and $Q_1$,$Q_2$ for $\Delta x=\lambda/2$). From Eq. \ref{eq:state}, we may verify that $|\psi_\textrm{ph}\rangle$ is a two-photon N00N state when the spatial separation between qubits is $\Delta x = \lambda/4, 3\lambda/4, ..., (2n+1)\lambda/4$, where $n$ is an integer. This can be understood by considering the interference between the four possible coherent emission pathways for two excitations to leave the system, shown in Fig.~\ref{fig:1}\textbf{B}. The emission pathways containing a single photon in both left and right propagating modes destructively interfere, resulting in the entangled state $|\psi_\textrm{ph}\rangle=(|20\rangle-|02\rangle)/\sqrt{2}$. Note that waveguide-mediated exchange interactions can be ignored because both qubits are fully excited. In contrast, for spacings $\Delta x = 0, \lambda/2, ...,  n\lambda/2$, depicted in Fig.~\ref{fig:1}\textbf{C}, the destructive interference no longer occurs, resulting in a standard (equal) partitioning of the photons into the left and right propagating modes. For this latter configuration, the decay of the qubits from $|ee\rangle$ to $|gg\rangle$ is determined by super-radiant emission \cite{VanLoo2013}.

Fig.~\ref{fig:2}\textbf{A} shows the control and measurement schematic. First, we measure the scattering of coherent microwave fields to extract qubit parameters and calibrate the absolute power of photons at the qubit (see supplementary info). Next, we independently prepare the qubits by detuning them from each other and then applying resonant microwave pulses to the transmission line. The qubits can be individually prepared anywhere on the Bloch sphere $\alpha|g\rangle +\beta |e\rangle$, where $\alpha$ and $\beta$ are complex coefficients determined by the amplitude and phase of the pulse. We then verify the state of the photons that are emitted by the qubits using quadrature amplitude detection of the left and right outputs of the transmission line. These photons are amplified and downconverted to an intermediate frequency $f_d$ using IQ mixing. For example, we can prepare a single detuned qubit in the state $(|g\rangle + |e\rangle)/\sqrt{2}$, which we use for calibration purposes (see below), and capture the time dynamics of the emission (Fig. \ref{fig:2}\textbf{B}) by averaging the voltage amplitudes $V_\textrm{L/R}^\textrm{I/Q}(t)$ at the output of the IQ mixers over many records. The qubit can also be fully excited to $|e\rangle$, as will be required for the N00N-state generation protocol. In this case, the emitted photon has no coherence relative to the vacuum state $|00\rangle$, and thus the voltage averages to zero as shown in Fig. \ref{fig:2}\textbf{C}

In order to uniquely identify the state and correlations of the photons emitted from two qubits, it is necessary to measure higher-order moments of the fields. To do this, time-independent values for the field quadratures of both the left $S_\textrm{L} = X_\textrm{L}\ +\ iP_\textrm{L}$ and right $S_\textrm{R} = X_\textrm{R}\ +\ iP_\textrm{R}$ emission signals are obtained through digital demodulation and integration of individual records of $V_\textrm{L/R}^\textrm{I/Q}(t)$. Using repeated measurements of these values, we construct a 4D probability distribution $Q(S_\textrm{L},S_\textrm{L}^*,S_\textrm{R},S_\textrm{R}^*)$ that are used to obtain the moments of $S_\textrm{L}$ and $S_\textrm{R}$ 
\begin{equation}
\label{eq:moments}
\begin{split}
&\langle {\hat{S}_\textrm{L}^{\dag w}} \hat{S}_\textrm{L}^x {\hat{S}_\textrm{R}^{\dag y}} \hat{S}_\textrm{R}^z \rangle = \\ &\int d^2S_\textrm{L} d^2S_\textrm{R}\ S_\textrm{L}^{*w} S_\textrm{L}^x S_\textrm{R}^{*y} S_\textrm{R}^z\ Q(S_\textrm{L},S_\textrm{L}^*,S_\textrm{R},S_\textrm{R}^*).
\end{split}
\end{equation}
We account for the noise added by the amplifiers in the measurement chain by using the input-output relations for a phase-insensitive amplifier $\hat{S}_\textrm{L(R)} = \sqrt{G_\textrm{L(R)}} \hat{a}_\textrm{L(R)} + \sqrt{G_\textrm{L(R)}-1}\hat{h}_\textrm{L(R)}^\dag$ \cite{Eichler2011,Caves1982,Eichler2012}, where $\hat{a}_\textrm{L(R)}$ is the left (right) output mode of the device, $\hat{h}_\textrm{L(R)}^\dag$ is the noise mode added by the left (right) amplification chain, and $G_\textrm{L(R)}$ is the gain of the left (right) amplification chain. The moments of the noise channels $\langle\hat{h}_\textrm{L}^w  {\hat{h}_\textrm{L}^{\dag x}} \hat{h}_\textrm{R}^y {\hat{h}_\textrm{R}^{\dag z}} \rangle$ are found by measuring the moments of $S_\textrm{L}$ and $S_\textrm{R}$ while leaving the qubits in the ground state. We account for residual thermal photons with an effective temperature $\approx \SI{46}{mK}$ in $\hat{a}_\textrm{L, R}$ when computing the statistics of the noise. The moments of the fields before amplification $\langle {\hat{a}_\textrm{L}^{\dag w}} \hat{a}_\textrm{L}^x {\hat{a}_\textrm{R}^{\dag y}} \hat{a}_\textrm{R}^z \rangle$ are determined by inverting the amplifier input-output relations (see supplementary info).

\begin{figure*}[t!]
    \centering
    \includegraphics[width=\textwidth]{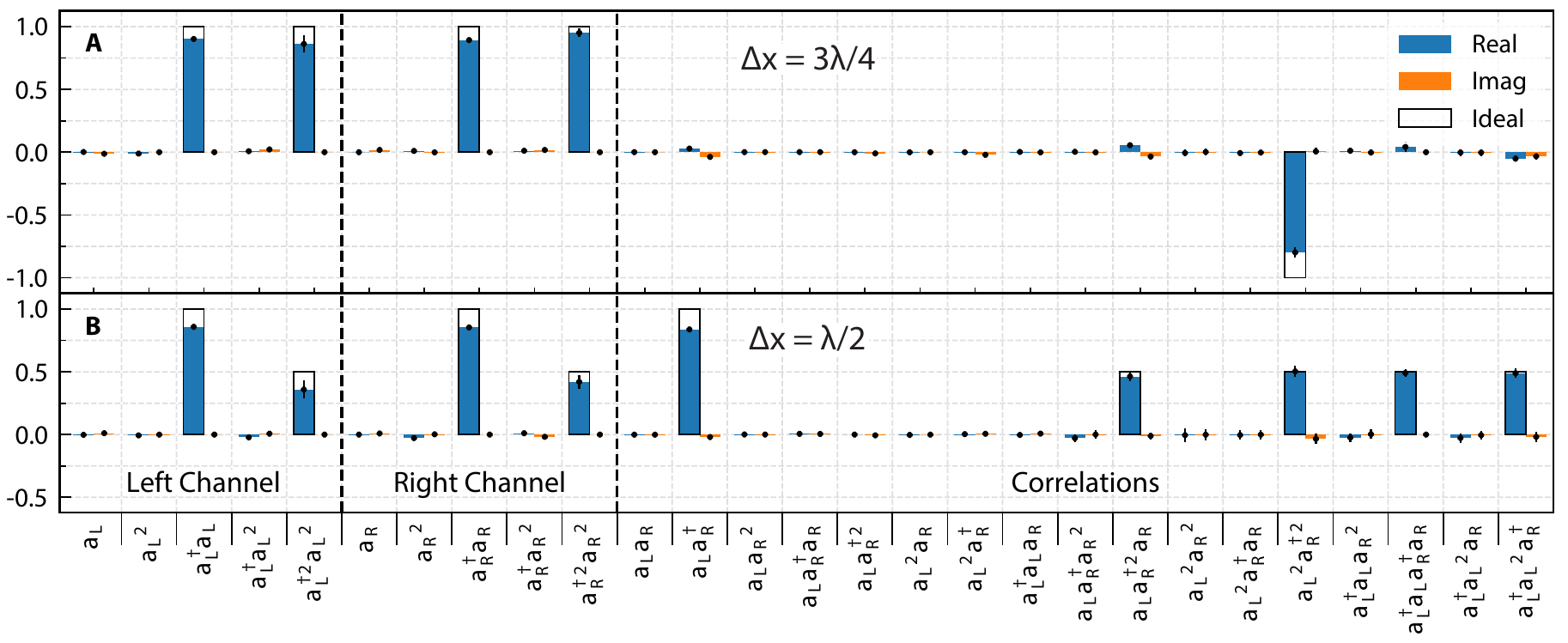}
    \caption{\textbf{Photon state tomography.}\ (\textbf{A}) Real (blue) and imaginary (orange) parts of the measured normally-ordered moments of the left- and right-propagating photonic fields $\langle {\hat{a}_\textrm{L}^{\dag w}} \hat{a}_\textrm{L}^x {\hat{a}_\textrm{R}^{\dag y}}  \hat{a}_\textrm{R}^z \rangle$ up to fourth order ($w, x, y, z \in \{0, 1, 2\}$)  for $\Delta x = 3\lambda/4$. The moments are separated according to their corresponding channel or correlations. The ideal values for the moments are given by the box frames around the measured values. (\textbf{B}) Measured and ideal moments for $\Delta x = \lambda/2$.   }
    \label{fig:moments}
\end{figure*}

Before generating the photonic states of interest, we first obtain the properties of the measurement chains. We are able to calibrate the net amplification gain by preparing a single qubit in an equal superposition of its ground and excited states \cite{Eichler2011}, as done in Fig. \ref{fig:2}\textbf{B}. For this case, the state of the emitted photon is $|00\rangle/\sqrt{2} + (|10\rangle + |01\rangle)/2$, since the photon is released symmetrically into both outputs of the waveguide. By taking advantage of the difference in scaling between first and second order moments with respect to $G_\textrm{L(R)}$, the gain can be calibrated by finding the value for which $\langle\hat{a}_\textrm{L(R)}\rangle = \sqrt{2} \langle\hat{a}_\textrm{L(R)}^\dag\hat{a}_\textrm{L(R)}\rangle$ is obtained from the inverted input-output relations of the amplifiers. Next, because the statistics of the noise modes are well-described by a thermal state $\hat{\rho}_h=\sum_i n_\textrm{noise}^i/(1+n_\textrm{noise})^{i+1} |i\rangle\langle i|$, where $n_\textrm{noise}$ is the average number of photons added by the noise, we can find the detection efficiency of our measurement chains $\eta = (1+n_\textrm{noise})^{-1}$ by performing a maximum-likelihood-estimation on the measured moments of $\hat{h}_\textrm{L, R}$. We extract the $n_\textrm{noise}$ that best describes the measurements and find the detection efficiencies to be $\eta_\textrm{L(R)} \approx 10.4\%\ (12.1\%)$. Finally, we alternate between initializing the two active qubits into the fully excited ($|ee\rangle$) and ground ($|gg\rangle$) states while measuring $\hat{S}_\textrm{L(R)}$ with a repetition period of $10 \mu$s to obtain the statistics of the emitted photons and the noise. 

We first initialize the qubits to $|\psi_\textrm{qb}\rangle = |ege\rangle$ with $Q_1$ and $Q_3$ separated by a distance of $\Delta x = 3\lambda/4$ along the waveguide. In doing so, we generate the two-photon N00N state $|\psi_\textrm{ph}\rangle = (|20\rangle - |02\rangle)/\sqrt{2}$ due to the complete destructive quantum interference of the $|11\rangle$ state, given by the phase factors shown in Fig.~\ref{fig:1}\textbf{B}. This is reminiscent of the final-state stimulation due to bosonic quantum statistics that is observed with identical photons in a Hong-Ou-Mandel experiment \cite{HOM,Lang2013}. We are able to validate the state of the emitted photons through the moments and correlations between the left and right output modes shown in Fig.~\ref{fig:moments}\textbf{A}. We observe $\langle \hat{a}_\textrm{L}^\dag \hat{a}_\textrm{L}\rangle \approx \langle \hat{a}_\textrm{R}^\dag \hat{a}_\textrm{R}\rangle \approx 1$, since there is one photon per mode on average. We also observe that the two-photon coincidences are $\langle {\hat{a}_\textrm{L}^{\dag 2}} {\hat{a}_\textrm{L}}^2\rangle \approx \langle {\hat{a}_\textrm{R}^{\dag 2}} {\hat{a}_\textrm{R}}^2\rangle \approx 1$, whereas the cross-coincidence is $\langle \hat{a}_\textrm{L}^\dag \hat{a}_\textrm{L} \hat{a}_\textrm{R}^\dag \hat{a}_\textrm{R} \rangle \approx 0$. These moments are consistent with two photons simultaneously arriving at the same detector rather than a single photon at each. Coherence between the $|20\rangle$ and $|02\rangle$ states is demonstrated via the two-photon cross-correlation: $\langle {\hat{a}_\textrm{L}}^2 {\hat{a}_\textrm{R}^{\dag 2}}\rangle \approx -1$.

\begin{figure*}[t!]

    \centering
    \includegraphics[]{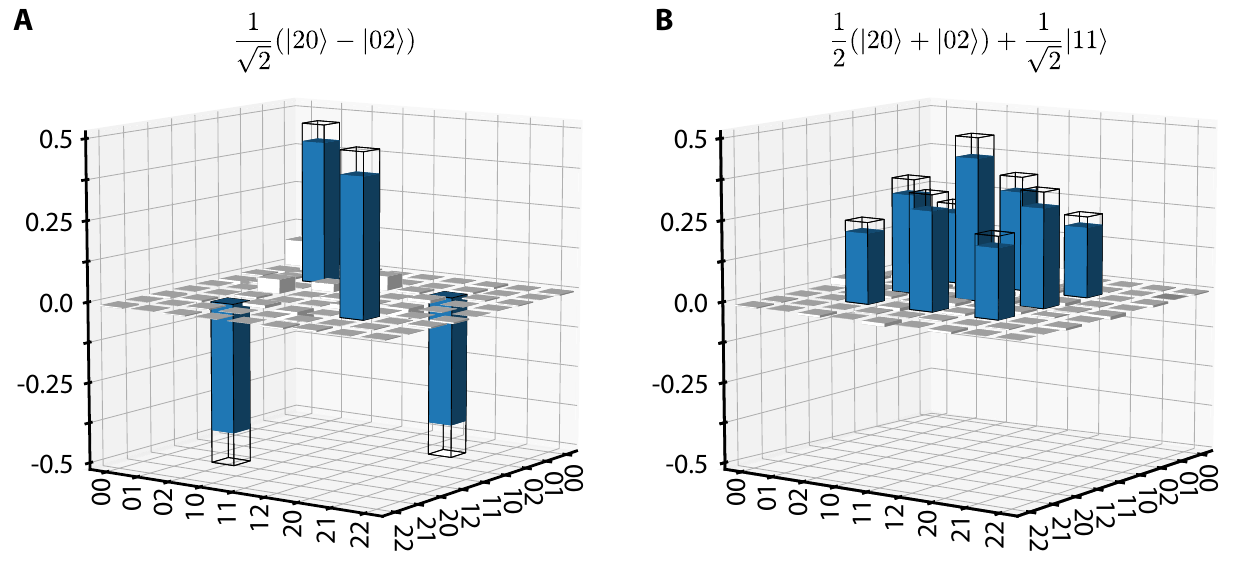}
    \caption{\textbf{Density matrix reconstruction of photonic states.}\ Real part of the density matrix in the Fock basis of the left and right propagating modes compared with expected state (wire frames) for (\textbf{A}) $(|20\rangle - |02\rangle )/\sqrt{2}$ at $\Delta x = 3\lambda/4$ and (\textbf{B}) $(|20\rangle + |02\rangle)/2 + |11\rangle/\sqrt{2}$ at $\Delta x = \lambda/2$. Density matrices are obtained via maximum-likelihood estimation on measured photonic moments with fidelities of 84$\%$ and 87$\%$, respectively. The matrix elements that are ideally non-zero are shaded in blue. The predominant source of infidelity is given by the finite population of 0.09 (A) and 0.11 (B) in the $|00\rangle\langle00|$ state. }
    \label{fig:dm}

\end{figure*}

We contrast the case of $\Delta x = 3\lambda/4$ with $\Delta x = \lambda/2$ to demonstrate the tunability of $|\psi_\textrm{ph}\rangle$. Here, we use $Q_1$ and $Q_2$ and initialize the qubits to $|\psi_\textrm{qb}\rangle = |eeg\rangle$. Constructive quantum interference of $|11\rangle$ leads to the output state $|\psi_\textrm{ph}\rangle = (|20\rangle + |02\rangle)/2 + |11\rangle/\sqrt{2}$ (Fig.~\ref{fig:1}\textbf{C}). The statistics of $|\psi_\textrm{ph}\rangle$ are now consistent with the standard partitioning of two classical particles, with each being independently and equally likely to appear in one of the two modes. The moments for this case are shown in Fig.~\ref{fig:moments}\textbf{B} and once again verify the predicted outcome. We obtain $\langle \hat{a}_\textrm{L}^\dag \hat{a}_\textrm{L}\rangle \approx \langle \hat{a}_\textrm{R}^\dag \hat{a}_\textrm{R}\rangle \approx 1$ as the average number of photons per mode remains unity. However, the two photons will now occupy the same mode only half of the time. As a result, two-photon coincidences $\langle {\hat{a}_\textrm{L}^{\dag 2}} {\hat{a}_\textrm{L}}^2\rangle \approx \langle {\hat{a}_\textrm{R}^{\dag 2}} {\hat{a}_\textrm{R}}^2\rangle \approx 1/2$ only occur $50\%$ of the time, compared to $100\%$ of the time for the two-photon N00N state. Additionally, we now observe a non-zero cross-coincidence $\langle \hat{a}_\textrm{L}^\dag \hat{a}_\textrm{L} \hat{a}_\textrm{R}^\dag \hat{a}_\textrm{R} \rangle \approx 0.5$, indicating that the photons arrive at opposite detectors the other $50\%$ of the time. Finally, the measurements of $\langle \hat{a}_\textrm{L} \hat{a}_\textrm{R}^\dag\rangle \approx 1$, $\langle {\hat{a}_\textrm{L}}^2 {\hat{a}_\textrm{R}^{\dag 2}}\rangle \approx 0.5$, and $\langle \hat{a}_\textrm{L}{\hat{a}_\textrm{R}^{\dag 2}}\hat{a}_\textrm{R}^\dag\rangle \approx \langle \hat{a}_\textrm{L}^\dag{\hat{a}_\textrm{L}}^2\hat{a}_\textrm{R}^\dag\rangle \approx 0.5$ demonstrate the appropriate coherences between the $|02\rangle$, $|20\rangle$, and $|11\rangle$ states.

To further characterize the state of the emitted photons, we obtain the density matrix $\hat{\rho}$ in the Fock-state basis by applying maximum-likelihood-estimation to the measured moments. The real part of $\hat{\rho}$ is shown in Fig.~\ref{fig:dm}, with the magnitude of all values in the imaginary part (not shown) being less than 0.037. The N00N state generated with $\Delta x = 3\lambda/4$ is clearly evident in Fig.~\ref{fig:dm}\textbf{A} with a trace overlap fidelity of $Tr(\hat{\rho}\hat{\sigma})=84$\%, where $\hat{\sigma}$ is the ideal density matrix. The density matrix for the emitted photons at $\Delta x = \lambda/2$ is shown in Fig.~\ref{fig:dm}\textbf{B} with a state preparation fidelity of 87\%. In both cases, we attribute a majority of the infidelity to waveguide-induced $T_1$ decay of the qubits during state initialization, as evidenced by a finite population of 0.09 and 0.11 in the $|00\rangle$ state of $\hat{\rho}$. Recent work has shown that this infidelity can be substantially reduced with the use of quantum interference with ``giant atoms'' \cite{FriskKockum2014,kannan2019}, where qubit-waveguide couplings can be tuned in-situ such that the qubits are not subject to waveguide-induced decoherence during state preparation. Furthermore, giant atoms can also be used to engineer tailored qubit-waveguide coupling, waveguide-mediated qubit-qubit coupling, and correlated decay spectra \cite{kannan2019} with the desired properties for a given interference condition.

Our results demonstrate that a wQED architecture supports high-fidelity generation of spatially entangled microwave photons. Our approach is extensible to higher-order photonic states through the addition of qubits, such that more photons are emitted, and with the appropriate choices of $\Delta x$ to obtain the desired quantum interference. These types of photonic states are also known to be useful for high-precision phase measurements in quantum metrology  \cite{dowling}. Although current limitations in detector efficiency hinder the ability to measure higher-order moments, and thus verify the resulting higher-order photonic states, recent proposals for number-resolved microwave photon detectors \cite{Royer2018,Kono2018} can address this issue. Finally, devices of the type studied in this work can be further generalized with the addition of direct qubit-qubit coupling, which can be used to dynamically select the direction in which photons are emitted or absorbed \cite{gheeraert2020bidirectional}. We envision an architecture where quantum information and entanglement are routed and spread throughout a quantum network via the quantum interference between the photons emitted by qubits that are coupled to a waveguide. Generating itinerant photons using the principles and techniques outlined in this work can then be applied towards realizing  interconnected quantum networks for both quantum communication and distributed quantum computation.



\section*{acknowledgments}
We would like to thank Jochen Braum\"uller and Antti Veps\"al\"ainen for valuable discussions.
This research was funded in part by the U.S. Department of Energy, Office of Science, Basic Energy Sciences, Materials Sciences and Engineering Division under Contract No. DE-AC02-05-CH11231 within the High-Coherence Multilayer Superconducting Structures for Large Scale Qubit Integration and Photonic Transduction program (QISLBNL);
and by the Department of Defense via MIT Lincoln Laboratory under U.S. Air Force Contract No. FA8721-05-C-0002.
B.K. gratefully acknowledges support from the National Defense Science and Engineering Graduate Fellowship program.
M.K. gratefully acknowledges support from the Carlsberg Foundation during a portion of this work.
The views and conclusions contained herein are those of the authors and should not be interpreted as necessarily representing the official policies or endorsements of the U.S. Government.

\bibliography{main}

\clearpage
\onecolumngrid
\setcounter{figure}{0}
\setcounter{equation}{0}
\setcounter{table}{0}
\renewcommand\theequation{S\arabic{equation}}
\renewcommand\thefigure{S\arabic{figure}}
\renewcommand\thetable{S\arabic{table}}

\renewcommand{\thesubsection}{\Alph{subsection}}
 \section*{Supplementary Information}
 \section{Experimental Setup}
 \begin{figure*}[h]
    \centering

    \includegraphics[width=\textwidth]{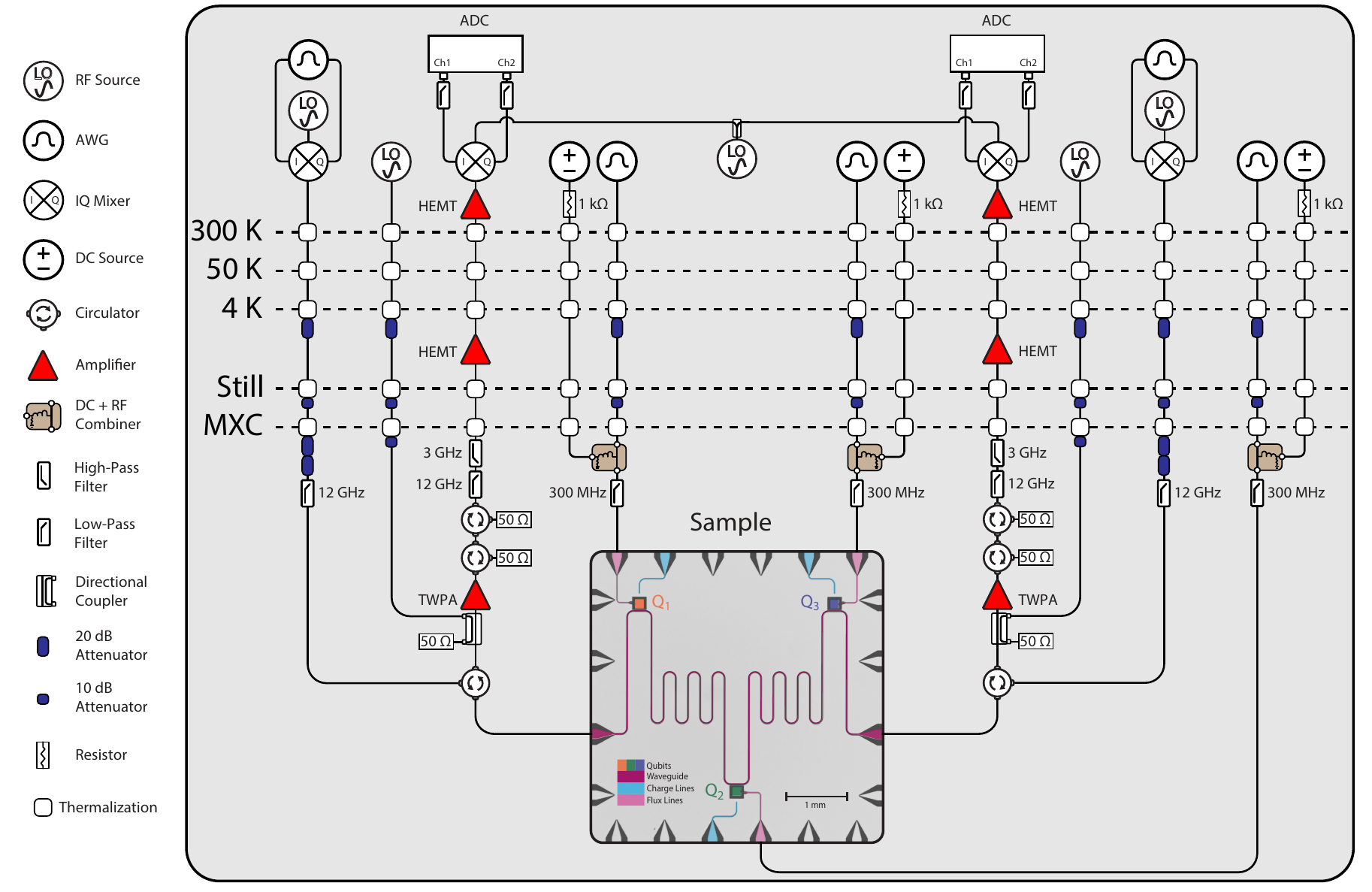}
    \caption{\textbf{Experimental Setup}. A schematic diagram of the experimental setup used to obtain the data presented in the main text. }
    \label{fig:expsetup} 
\end{figure*}
The experiments are performed in a Bluefors XLD600 dilution refrigerator, capable of cooling to a base temperature of $\SI{10}{mK}$. The sample is placed between two circulators for double-sided input and output. Both inputs are attenuated by 20dB at the $\SI{4}{K}$ stage, 10dB at the still, and 40dB at the mixing chamber (MXC) to ensure proper thermalization of the line. The samples are magnetically shielded at the MXC by superconducting and Cryoperm-10 shields. A Josephson travelling wave parametric amplifier (TWPA) \cite{Macklin2015} is used as the first amplifier in the measurement chain. The TWPAs are pumped in the forward direction using a directional coupler. The readout signal is filtered with $\SI{3}{GHz}$ high-pass and $\SI{12}{GHz}$ low-pass filters. Two additional circulators are placed after the TWPA in the MXC to prevent noise from higher-temperature stages travelling back into the TWPA and the sample. High electron mobility transistor (HEMT) amplifiers are used at $\SI{4}{K}$ and room-temperature stages of the measurement chain for further amplification. The signal is then downconverted to an intermediate frequency using an IQ mixer, filtered, digitized, and demodulated with custom FPGA code. 

The frequencies of the qubits are controlled with local flux lines. Each flux line has both DC and RF control that are combined and filtered with $\SI{300}{MHz}$ low-pass filters at the mixing chamber. The RF flux control line is attenuated by 20dB at the $\SI{4}{K}$ stage and by 10dB at the still. A $\SI{1}{k\Omega}$ resistor is placed in series with the DC voltage source to generate a DC current. Although the chip has individual charge lines for each qubit, all qubit drives and initialization pulses are applied via the central transmission line.

\section{Spectroscopic Measurements}
 \begin{figure*}[h]
    \centering

    \includegraphics[width=\textwidth]{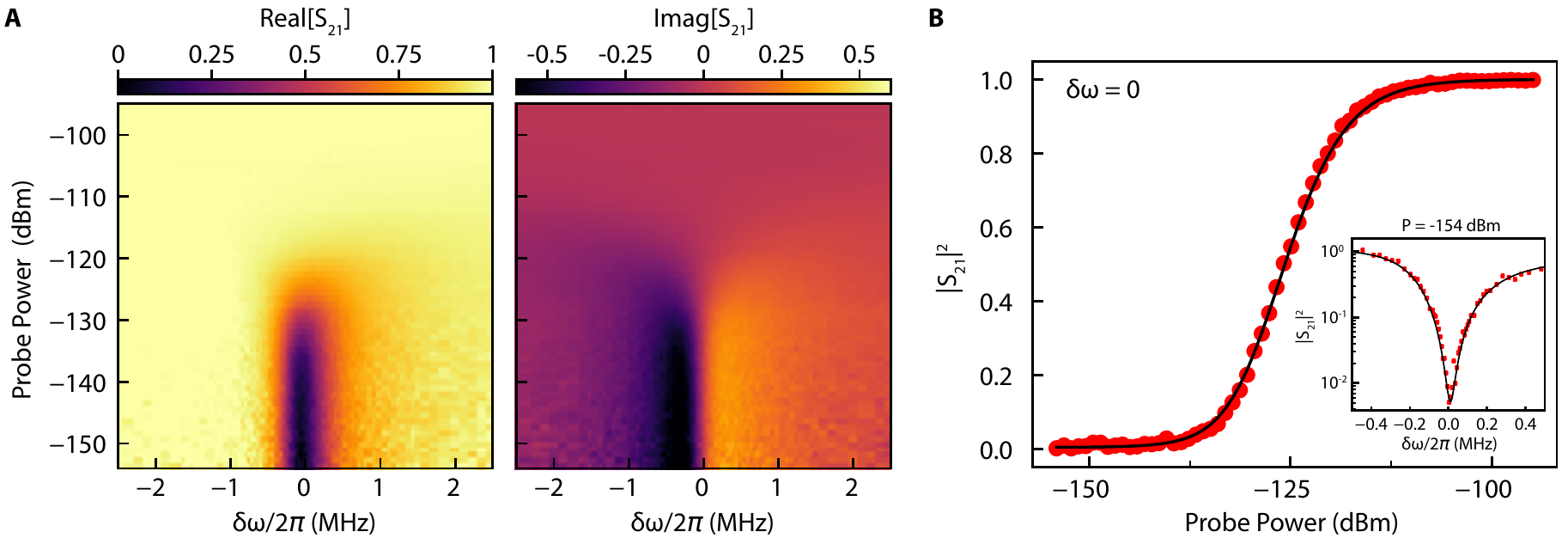}
    \caption{\textbf{Qubit Spectroscopy}. (\textbf{A}) Real (left panel) and imaginary (right panel) components of the transmission $S_{21}$ of a coherent probe tone as a function of qubit-probe detuning $\delta \omega/2\pi$ and the probe power. (\textbf{B}) Transmittance $|S_{21}|^2$ at zero detuning $\delta \omega = 0$ as a function of probe power. The theoretical fit (black line) is plotted over the measured data (red dots). A trace of the qubit’s frequency response at a probe power $P=\SI{-154}{dBm}$ is shown in the inset (bottom right). }
    \label{fig:spectroscopy} 
\end{figure*}

Individual qubits that are coupled to a waveguide will act as single-photon mirrors \cite{Hoi2011}. The non-linearity of the qubit causes the transmittance (and reflectance) of coherent probe tones incident upon the qubit to depend on the probes power  \cite{Astafiev2010,Hoi2011,Hoi2013}. This phenomenon can be derived by modelling the system with a master equation resultant from total system-bath Hamiltonian of Eq. \ref{eq:hamiltonian} \cite{Lalumiere2013,Mirhosseini2019}. For a single qubit, the master equation reduces to

\begin{equation}
    \hat{\rho} = \frac{i}{\hbar} \left[\hat{H},\hat{\rho}\right] + (1+n_\textrm{th})\gamma \mathcal{D}\left[\hat{\sigma}_-\right] \hat{\rho} + n_\textrm{th}\gamma \mathcal{D}\left[\hat{\sigma}_+\right] + \frac{\gamma_\phi}{2} \mathcal{D}\left[\hat{\sigma}_z\right]
\end{equation}

\begin{equation}
    \hat{H} = \frac{1}{2} \delta \omega \hat{\sigma}_z + \frac{1}{2} \Omega_p \hat{\sigma}_x,
\end{equation}
where $\mathcal{D}\left[O\right] \rho = O\rho O^\dag - \frac{1}{2} \{O^\dag O, \rho\}$ is the standard Lindblad dissipator, $n_\textrm{th} = (e^{\hbar \omega / k_B T} - 1)^{-1}$ is the average number of thermal photons in the bath at a temperature $T$, $\gamma_\phi$ is the qubit dephasing rate, $\delta \omega = \omega - \omega_p$ is the qubit-probe detuning,  and $\Omega_p = \sqrt{2\gamma P/\hbar \omega}$ is the strength of the coherent probe with power $P$. Using the input-output relations of Eq. \ref{eq:inout}, the complex transmission amplitude $S_{21}=\langle \hat{a}_\textrm{L/R}\rangle/ \langle \hat{a}_\textrm{L/R}^\textrm{in} \rangle$ is \cite{Mirhosseini2019}

\begin{equation}
\label{eq:s21}
    S_{21}(\delta \omega, \Omega_p) = 1 - \frac{\gamma (1 - i\frac{\delta \omega}{\gamma_2})}{2\gamma_2(1+2n_\textrm{th}) \left[1 + \left(\frac{\delta \omega}{\gamma_2}\right)^2 + \frac{\Omega_p^2}{(1+2n_\textrm{th}) \gamma \gamma_2}\right]},
\end{equation}
where $\gamma_2 = (1+2n_\textrm{th}) \gamma/2 + \gamma_\phi$ is the total decoherence rate of the qubit. This measurement allows us to extract the qubit parameters, and its non-linearity enables the calibration of the absolute power of photons at the qubit. Fig. \ref{fig:spectroscopy}A plots the real and imaginary parts the measured $S_{21}$ as a function of both qubit-probe detuning $\delta \omega$ and probe power $P$. We perform a joint 2D fit of Eq. \ref{eq:s21} on this data to extract the qubit parameters: $\gamma/2\pi \approx \SI{0.53}{MHz}$, $\gamma_\phi/2\pi \approx \SI{51}{kHz}$, and $n_\textrm{th} \approx 0.006$, which at a frequency of 4.85 GHz corresponds to an effective temperature of 46 mK. The measured transmittance $|S_{21}|^2$ at zero detuning $\delta \omega = 0$ is plotted with the theoretical fit in Fig. \ref{fig:spectroscopy}B.

\section{Moment Inversion}

We describe an efficient procedure for determining the moments of the field before amplification, $\langle \hat{a}^{\dag n}_\textrm{L} \hat{a}^m_\textrm{L} \hat{a}^{\dag k}_\textrm{R} \hat{a}^l_\textrm{R} \rangle$, where $n, m ,k ,l\in \{0,N\}$ are integers up to a desired moment order $N$. In our experiment, we consider moments of order up to $N=2$. The standard input-output relationship for phase insensitive amplifiers is given by
\begin{equation}
    \hat{S}_\textrm{L(R)}= \sqrt{G_\textrm{L(R)}}\hat{a}_\textrm{L(R)}+ \sqrt{G_\textrm{L(R)}-1}\hat{h}_\textrm{L(R)}^\dag,
    \label{eqn:long}
\end{equation}
where $\hat{a}_\textrm{L(R)}$ is the left (right) output mode of the device, $\hat{h}_\textrm{L(R)}^\dag$ is the noise mode added by the left (right) amplification chain, and $G_\textrm{L(R)}$ is the amplification of the left (right) amplification chain~\cite{Eichler2011,Caves1982,Eichler2012}. When the gain of the amplifiers are large ($G_\textrm{L(R)} \gg 1$), as is the case in our setup, Eq.~\ref{eqn:long} can be simplified to
\begin{equation}
        \hat{S'}_\textrm{L(R)} = \frac{\hat{S}_\textrm{L(R)}}{\sqrt{G_\textrm{L(R)}}}\approx\hat{a}_\textrm{L(R)}+\hat{h}_\textrm{L(R)}^\dag.
\end{equation}
Furthermore, we assume that modes of interest $\hat{a}_\textrm{L(R)}$ and noise modes $\hat{h}_\textrm{L(R)}$ are uncorrelated. Under these conditions, the moments of $\hat{S'}_\textrm{L(R)}$, $\hat{a}_\textrm{L(R)}$, and $\hat{h}_\textrm{L(R)}^\dag$ are related as
\begin{equation}
\label{eqn:analytic}
\begin{split}
    & \langle {\hat{S'}_\textrm{L}^{\dag n}} \hat{S'}_\textrm{L}^m {\hat{S'}_\textrm{R}^{\dag k}} \hat{S'}_\textrm{R}^l \rangle=
    \sum_{w=0}^n \sum_{x=0}^m \sum_{y=0}^k \sum_{z=0}^l
    {n \choose w} {m \choose x} {k \choose y} {l \choose z}
    \langle \hat{a}^{\dag w}_\textrm{L} \hat{a}^x_\textrm{L} \hat{a}^{\dag y}_\textrm{R} \hat{a}^z_\textrm{R} \rangle
    \langle\hat{h}_\textrm{L}^{n-w}  {\hat{h}_\textrm{L}^{\dag m-x}} \hat{h}_\textrm{R}^{k-y} {\hat{h}_\textrm{R}^{\dag l-z}} \rangle.
\end{split}  
\end{equation}
As described in the main text, we use heterodyne detection on the output of the measurement chain to form a 4D probability distribution, $Q(S_\textrm{L}, S_\textrm{L}^*, S_\textrm{R}, S_\textrm{R}^*)$, from which the moments of $S'_\textrm{L}$ and $S'_\textrm{R}$ can be obtained,
\begin{equation}
\label{eq:moments}
\langle {\hat{S'}_\textrm{L}^{\dag n}} \hat{S'}_\textrm{L}^m {\hat{S'}_\textrm{R}^{\dag k}} \hat{S'}_\textrm{R}^l \rangle = \int d^2S_\textrm{L} d^2S_\textrm{R}\ S_\textrm{L}^{*n} S_\textrm{L}^m S_\textrm{R}^{*k} S_\textrm{R}^l\ Q(S_\textrm{L},S_\textrm{L}^*,S_\textrm{R},S_\textrm{R}^*) G_\textrm{L}^{-\frac{n+m}{2}}
    G_\textrm{R}^{-\frac{k+l}{2}}.
\end{equation}
To obtain the moments of the noise added by the amplifiers  $\langle\hat{h}_\textrm{L}^{n}  {\hat{h}_\textrm{L}^{\dag m}} \hat{h}_\textrm{R}^{k} {\hat{h}_\textrm{R}^{\dag l}} \rangle$, the qubits are left in their ground states. If the temperature of $\hat{a}_\textrm{L/R}$ is small $k_BT \ll \hbar \omega$, then the state of these photonic modes can be approximated as vacuum. Under this condition, we have
\begin{equation}
  \langle \hat{a}^{\dag w}_\textrm{L} \hat{a}^x_\textrm{L} \hat{a}^{\dag y}_\textrm{R} \hat{a}^z_\textrm{R} \rangle =
    \begin{cases}
      1 & \text{if $w,x,y,z=0$}\\
      0 & \text{otherwise}
    \end{cases}.      
\end{equation}
Eq.~\ref{eqn:analytic} is then significantly reduced such that moments of the noise channels can be directly obtained from the measured moments of $\langle {\hat{S'}_\textrm{L}^{\dag n}} \hat{S'}_\textrm{L}^m {\hat{S'}_\textrm{R}^{\dag k}} \hat{S'}_\textrm{R}^l \rangle_0$ with $|\psi_\textrm{ph}\rangle=|00\rangle$
\begin{equation}
    \langle\hat{h}_\textrm{L}^{n}  {\hat{h}_\textrm{L}^{\dag m}} \hat{h}_\textrm{R}^{k} {\hat{h}_\textrm{R}^{\dag l}} \rangle = \langle {\hat{S'}_\textrm{L}^{\dag n}} \hat{S'}_\textrm{L}^m {\hat{S'}_\textrm{R}^{\dag k}} \hat{S'}_\textrm{R}^l \rangle_0.
\end{equation}
After determining both $\langle {\hat{S'}_\textrm{L}^{\dag n}} \hat{S'}_\textrm{L}^m {\hat{S'}_\textrm{R}^{\dag k}} \hat{S'}_\textrm{R}^l \rangle$ and $\langle\hat{h}_\textrm{L}^n  {\hat{h}_\textrm{L}^{\dag m}} \hat{h}_\textrm{R}^k {\hat{h}_\textrm{R}^{\dag l}} \rangle$, we can solve for $\langle \hat{a}^{\dag w}_\textrm{L} \hat{a}^x_\textrm{L} \hat{a}^{\dag y}_\textrm{R} \hat{a}^z_\textrm{R} \rangle$ by inverting a system of linear equations. We begin by defining vectors $\vec{S}$ and $\vec{a}$, where the elements are all possible combinations of $\langle {\hat{S'}_\textrm{L}^{\dag n}} \hat{S'}_\textrm{L}^m {\hat{S'}_\textrm{R}^{\dag k}} \hat{S'}_\textrm{R}^l \rangle$ and $\langle
\hat{a}^{\dag w}_\textrm{L} \hat{a}^x_\textrm{L} \hat{a}^{\dag y}_\textrm{R} \hat{a}^z_\textrm{R} \rangle$, respectively. These vectors are length $(N+1)^4$ and takes the form,
\begin{equation}
    \vec{S}=
    \begin{bmatrix}
    1\\
    \langle \hat{S'}_\textrm{R} \rangle \\
    \langle \hat{S'}_\textrm{R}^2 \rangle \\
    \vdots \\
    \langle \hat{S'}_\textrm{R}^N \rangle \\
    \langle \hat{S'}_\textrm{R}^\dag \rangle \\
    \langle \hat{S'}_\textrm{R}^\dag\hat{S'}_\textrm{R} \rangle \\
    \vdots \\
    \langle \hat{S'}_\textrm{R}^{\dag N}\hat{S'}_\textrm{R}^N \rangle \\
    \vdots \\
    \langle \hat{S'}_\textrm{L}^{N} \hat{S'}_\textrm{R}^{\dag N} \hat{S'}_\textrm{R}^{N} \rangle \\
    \vdots \\
    \langle \hat{S'}_\textrm{L}^{\dag N} \hat{S'}_\textrm{L}^{N} \hat{S'}_\textrm{R}^{\dag N} \hat{S'}_\textrm{R}^{N} \rangle
    \end{bmatrix}\; \;,\; \; 
    \vec{a}=
    \begin{bmatrix}
    1\\
    \langle \hat{a}_\textrm{R} \rangle \\
    \langle \hat{a}_\textrm{R}^2 \rangle \\
    \vdots \\
    \langle \hat{a}_\textrm{R}^N \rangle \\
    \langle \hat{a}_\textrm{R}^\dag \rangle \\
    \langle \hat{a}_\textrm{R}^\dag\hat{a}_\textrm{R} \rangle \\
    \vdots \\
    \langle \hat{a}_\textrm{R}^{\dag N}\hat{a}_\textrm{R}^N \rangle \\
    \vdots \\
    \langle \hat{a}_\textrm{L}^{N} \hat{a}_\textrm{R}^{\dag N} \hat{a}_\textrm{R}^{N} \rangle \\
    \vdots \\
    \langle \hat{a}_\textrm{L}^{\dag N} \hat{a}_\textrm{L}^{N} \hat{a}_\textrm{R}^{\dag N} \hat{a}_\textrm{R}^{N} \rangle
    \end{bmatrix}.
\end{equation}
We can then relate $\vec{a}$ to $\vec{S}$ by a matrix $\pmb{H}$, such that $\vec{S}=\pmb{H}\vec{a}$. This matrix will have dimensions $(N+1)^4 \times (N+1)^4$, and be of the form

\begin{equation}
        \pmb{H}=
        \begin{bmatrix}
        1 & 0 & 0 &\cdots& 0& 0& 0 &\cdots &0 \\
        \langle \hat{h}_\textrm{R}^{\dag}\rangle & 1  & 0 & \cdots & 0 & 0 & 0 & \cdots & 0 \\
        \langle \hat{h}_\textrm{R}^{\dag 2}\rangle & 2\langle \hat{h}_\textrm{R}^{\dag}\rangle & 1 &\cdots& 0& 0& 0 &\cdots &0 \\
        \vdots & \vdots & \vdots & \ddots & \vdots & \vdots & \vdots & \ddots & \vdots \\
        \langle \hat{h}_\textrm{R}^{\dag N}\rangle & {N \choose 1}\langle \hat{h}_\textrm{R}^{\dag N-1}\rangle & {N \choose 2}\langle \hat{h}_\textrm{R}^{\dag N-2}\rangle & \cdots& 1& 0& 0 &\cdots &0 \\
        \langle \hat{h}_\textrm{R}\rangle & 0 & 0 & \cdots & 0 & 1 & 0 &\cdots &0 \\
        \langle \hat{h}_\textrm{R}\hat{h}_\textrm{R}^{\dag}\rangle & \langle \hat{h}_\textrm{R}\rangle & 0 & \cdots & 0 & \langle \hat{h}_\textrm{R}^{\dag}\rangle& 1 &\cdots &0 \\
        \vdots & \vdots & \vdots & \ddots & \vdots & \vdots & \vdots & \ddots & \vdots\\
        \langle \hat{h}_\textrm{L}^N\hat{h}_\textrm{L}^{\dag N}\hat{h}_\textrm{R}^N\hat{h}_\textrm{R}^{\dag N}\rangle & {N \choose 1}\langle \hat{h}_\textrm{L}^N\hat{h}_\textrm{L}^{\dag N}\hat{h}_\textrm{R}^N\hat{h}_\textrm{R}^{\dag N-1}\rangle & \cdots & \cdots& \langle \hat{h}_\textrm{L}^N\hat{h}_\textrm{L}^{\dag N}\hat{h}_\textrm{R}^N\rangle& {N \choose 1}\langle \hat{h}_\textrm{L}^N\hat{h}_\textrm{L}^{\dag N}\hat{h}_\textrm{R}^{N-1}\hat{h}_\textrm{R}^{\dag N}\rangle& \cdots &\cdots &1\\
        \end{bmatrix},
\end{equation}
The moments in the $\vec{a}$ can then be solved for by inverting $\pmb{H}$: $\vec{a}=\pmb{H}^{-1}\vec{S}$. Note that the matrix $\pmb{H}$ is lower-triangular, and thus the system can be solved efficiently using back-substitution.
\end{document}